\documentclass[usenatbib]{mn2e}

\usepackage{verbatim}
\usepackage{graphicx}
\usepackage{url}
\usepackage{amsfonts}
\usepackage{amsmath,amssymb}

\newcommand{\lb}[0] { \left( }
\newcommand{\rb}[0] { \right) }
\newcommand{\g}[0] { \gamma }

\newcommand{\beqs} { \begin{eqnarray} }
\newcommand{\eeqs} { \end{eqnarray} }
\newcommand{\bsub}{ \begin{subequations} }
\newcommand{\esub}{ \end{subequations} }
\newcommand{\nn} {\nonumber}

\renewcommand{\eqref}[1]{(\ref{#1})}
\newcommand{\mE} {\mathcal{E}}

\newcommand{\ergscm}[0] {\textrm{ erg s$^{-1}$ cm$^{-3}$}}

\newcommand{\erg}[0] {\textrm{ erg}}
\newcommand{\MeV}[0] {\textrm{ MeV}}
\newcommand{\Kel}[0] {\textrm{ K}}
\newcommand{\cm}[0] {\textrm{ cm}}

\newcommand{\EE}[2] {#1 \times 10^{#2}}

\title{The effect of neutrinos on the initial fireballs in gamma-ray bursts}
\author[Hylke B.J. Koers and Ralph A.M.J. Wijers]{Hylke B.J. Koers$^{1,2}$\thanks{E-mail:
hkoers@nikhef.nl} and Ralph A.M.J. Wijers$^{3}$\thanks{E-mail:
rwijers@science.uva.nl} \\
$^{1}$NIKHEF, P.O. Box 41882, 1009 DB Amsterdam, The Netherlands \\
$^{2}$University of Amsterdam, Amsterdam, The Netherlands \\
$^{3}$Anton Pannekoek Instituut,  Kruislaan 403, 1098 SJ Amsterdam, The Netherlands \\
\\
\emph{NIKHEF-2005-008} \\
\emph{astro-ph/0505533}}

\begin{document}
\pagerange{\pageref{firstpage}--\pageref{lastpage}} \pubyear{2005}

\maketitle

\label{firstpage}

\begin{abstract} 
We investigate the fate of very compact, sudden energy depositions 
that may lie at the origin of gamma-ray bursts. Following on from the
work of Cavallo and Rees (1978), we take account of the much higher
energies now believed to be involved. The main effect of this is that
thermal neutrinos are present and energetically important. We show that
these may provide sufficient cooling to tap most of the explosion energy.
However, at the extreme energies usually invoked for gamma-ray bursts,
the neutrino opacity suffices to prevent dramatic losses, provided that
the heating process is sufficiently fast. In a generic case, a few tens
of percent of the initial fireball energy will escape as an isotropic
millisecond burst
of thermal neutrinos with a temperature of about 60\,MeV, which is
detectable for nearby gamma-ray bursts and hypernovae. For parameters
we find most likely for gamma-ray burst fireballs, the dominant
processes are purely leptonic, and thus the baryon loading of the
fireball does not affect our conclusions.
\end{abstract}


\section{Introduction}

Due to their tremendous energy, and in view of the connections
discovered in recent years between gamma-ray bursts and massive stars
(e.g., \cite{Paradijs2000}, and references therein), it is now generally assumed that a gamma-ray burst (GRB) is initiated
when a few solar masses of material collapse to near their Schwarzschild 
radius. In the simplest possible models of what happens next, a fair
fraction of the gravitational energy released in the collapse is deposited
into a volume somewhat larger than that of the horizon of the collapsed
mass. The subsequent evolution of such a volume of highly concentrated
energy -- termed `fireball' -- was explored by Cavallo and Rees (\citeyear{1978MNRAS.183..359C}). These authors introduced a compactness parameter
for the volume, which expresses how easily a plasma consisting of 
baryons, photons, electrons and positrons can emit energy within a
dynamical time. For small compactness, the emission is easy and the
fireball cools by radiation. For large compactness, photons are trapped
and cooling occurs by adiabatic expansion: an explosion results in which
a significant fraction of the initial fireball energy is converted to
bulk kinetic energy of a relativistic outflow, a condition now thought
necessary for producing a gamma-ray burst.

At the time, Cavallo and Rees considered still relatively nearby origins
of GRBs, for which the required fireball energies imply conditions that
justified their assumption for the fireball composition. However, with
cosmological distances to GRBs the required fireball energies are now so
large that conditions of copious neutrino production become quite plausible.
Motivated by the concern that these neutrinos easily leave their creation
site due to their weak interaction with matter, and thereby carry away enough
energy to weaken or prevent an explosion, we investigate the evolution
of neutrino-rich fireballs. Neutrino emission was previously considered
as a sink of fireball energy, e.g.\ by Kumar (\citeyear{Kumar1999}), who
included emission of neutrinos in the optically thin limit. Neutrino
emissivity has been more widely studied in a slightly different context,
namely the evolution
post-collapse of the accretion disk or torus around the newborn black
hole, which may tap the accretion energy of the torus to power a GRB
\citep{Woosley1993}. The effect of neutrino opacity in this process
has been the subject of a few recent studies, e.g., by Lee et ~al.\
(\citeyear{Lee2004}) and by Janiuk et~al.\ (\citeyear{Janiuk2004}).

Here we study the evolution of a spherical fireball with given initial
radius, energy, and baryon content. We aim to be general in the 
physical processes we consider, but accept a few a priori constraints
on the initial parameters of the fireball: its initial energy must suffice
to power a GRB, hence it should be within a few decades of $10^{52}$\,erg;
its initial size cannot be much larger than the Schwarzschild radius of a
few solar masses, say $10^{6.5}$\,cm, because the mass must collapse to
such small radii in order to liberate such a large energy. Lastly, the
initial ratio of fireball energy to rest mass, $M_0$, of the entrained
baryons, $\eta\equiv\mathcal{E}/M_0c^2$, must be several hundred (corresponding
to almost 1\,TeV/baryon) in order
that eventually the baryons may be accelerated to a Lorentz factor 
high enough to produce a GRB. This combination of constraints implies that
the fireballs we study here are always very compact in the Cavallo and
Rees (or electromagnetic) sense. It also implies, as we show here, that
the baryons are relatively unimportant in the neutrino processes.

This paper is organized as follows: in section \ref{sect:general},
we discuss some general properties of the fireball environment.
Using these, we investigate the most important neutrino interactions
in this environment in section \ref{sect:neutrinos}. We introduce
the emissivity parameter $\chi$ and the optical depth $\tau$ to
describe the neutrino physics, and we draw a phase diagram for the
neutrino fireball. The dynamical evolution of the neutrino fireball
is discussed in section \ref{sect:evol}. The neutrino emission
is discussed in section \ref{sect:nuemission} and we present our
conclusions in section \ref{sect:conclusions}.

\section{General properties}
\label{sect:general}

\subsection{Composition and temperature}
The term `fireball' refers to a plasma consisting of photons, electrons and positrons, possibly with a small baryonic load \citep{1978MNRAS.183..359C}.
In this paper, we extend this to fireballs that contain neutrinos\footnote{Unless the difference is important, we will use the word `neutrinos' if we mean `neutrinos and antineutrinos'.}.
We consider a fireball that is initially opaque to neutrinos of all flavours. At some point during the fireball's expansion
(to be discussed in section \ref{sect:phasandevol}),  it becomes transparent
to muon- and tau-neutrinos, that subsequently decouple from the plasma. The electron-neutrinos decouple a bit later, which divides the plasma parameter space in three regions: region I where the fireball contains neutrinos of all flavours; region II where it contains only electron-neutrinos; and region III where all the neutrinos are decoupled. 

In thermodynamic equilibrium, the energy density and temperature are related by
\beqs
\frac{\mE}{V} =  g a  T^4 \, , 
\eeqs
where $a$ is the radiation constant, and  $g$ is a prefactor that depends on
the composition of the system. For the three regions introduced above:
\beqs
g_{\textrm{I}} = \frac{43}{8} \, , \qquad
g_{\textrm{II}}= \frac{29}{8} \, , \qquad
g_{\textrm{III}} = \frac{22}{8} \, .
\eeqs
Assuming a spherical configuration, the temperature of the plasma can be   expressed in terms of the energy and radius as
\beqs
\label{eq:temp}
\lb T_{11} \rb^4 = \frac{100}{g} \lb \mE_{52} \rb \lb R_{6.5} \rb^{-3} \, ,
\eeqs
where $T = T_{11}  \times 10^{11} \Kel $, $ \mE  =  \mE_{52} \times 10^{52} \erg$ and
\mbox{$R  =  R_{6.5}  \times  10^{6.5} \cm$}.

We use the following values for the initial fireball energy and radius as a reference (denoted with an asterisk):
\begin{subequations}
\label{eq:refs}
\beqs
\mE_* & =  & 10^{52} \erg \, , \\
R_* & = &  10^{6.5} \cm \, .
\eeqs
The reference temperature is
\beqs
T_* = \EE{2.1}{11} \Kel  = 17.9 /k_B \MeV \, .
\eeqs
\end{subequations}

\subsection{Baryons}
As the temperature is higher than typical binding energies,
nuclei are dissociated into nucleons. Hence `baryons' means `nucleons' in what follows (`baryon' is
however the standard terminology).
The requirement that  there should be \mbox{1 TeV} of energy available for the baryons
leads to a maximum number density of
\beqs
\label{eq:ref:nb}
n_{B,*} =  \EE{4.7}{31} \cm^{-3} \, ,
\eeqs
which will be used as the reference value in this study. 
It implies a baryonic mass density of $\rho_{B,*} = \EE{9.4}{7} \textrm{ gr cm}^{-3} $, which corresponds to a 
total baryonic mass of $\EE{6.2}{-6}$ solar masses contained in the volume $V_*$.
Note that the
nucleons are non-degenerate.

Because of overall charge neutrality, the ratio of protons to
neutrons can be expressed in terms of the electron fraction $Y_e$:
\bsub
\beqs
n_B & = & n_n + n_p \, , \\
 n_p & = &  Y_e n_B  =  \Delta n_e  \, ,
\eeqs
\esub
where \mbox{$\Delta n_e =  n_{e^-} - n_{e^+}$} is the net electron density.
The exact value of $Y_e$ in a physical situation is determined by beta-equilibrium conditions; see
e.g. \citet{Yuan:2003cg} and \citet{Beloborodov:2002af}. We will see in the next
section that the exact value of $Y_e$ is not very important for our purposes.

\subsection{Electron and positron number densities}
\label{sect:numdens}
Since $T_* \gg m_e c^2$, the electrons and positrons are extremely relativistic.
Using $E =pc$, the net electron density and
the combined electron-positron density \mbox{$n_e  =   n_{e^-} + n_{e^+}$} can be expressed as
\bsub
\beqs
\label{eq:thermo:deltanep} \Delta n_e  & = &  \frac{1}{3  \hbar^3 c^3} \lb (k_B T)^2 \mu_e + \frac{{\mu_e}^3}{\pi^2} \rb \, , \\
\label{eq:thermo:nep}
  n_e  & = & 0.37 \frac{ (k_B T)^3 }{\hbar^3 c^3} + \mathcal{O} \lb \mu_e \rb^2 \, ,
\eeqs
\esub
where $\mu_e$ is the electron chemical potential.

By definition, $Y_e < 1 $, so that $\Delta n_e \leq n_B$.
This places an upper bound on the net electron density and, through
eq. \eqref{eq:thermo:deltanep}, on the
electron chemical potential.
With the reference  baryon number density  of eq. \eqref{eq:ref:nb}, we find that the electron chemical potential is very small:
$\mu_e / (k_B T_* ) \sim \EE{2}{-4} \ll 1 $. 
From \eqref{eq:thermo:nep}, neglecting the chemical potential, we
find that at the reference temperature $T_*$:
\beqs
n_{e^-,*} = n_{e^+,*} = \EE{1.4}{35} \cm^{-3} \, .
\eeqs
Concluding, the fireball under consideration here is
nucleon-poor ($n_B \ll n_e$) and has a very small electron chemical 
potential ($\Delta n_e \ll n_e$).
This implies that the electrons and positrons are non-degenerate.

\section{Fireball neutrino physics}
\label{sect:neutrinos}

\subsection{The dominant neutrino processes}
The relative importance of interactions 
between neutrinos and the other components of the plasma depends on the temperature, the electron
chemical potential and the baryon density. The most important neutrino
production processes are discussed  in appendix \ref{sect:appQ}. Scattering and absorption processes are
discussed in appendix \ref{sect:appmfp}. We use the fact
that nucleons, electrons and positrons are non-degenerate.

For the present baryon densities, we observe from 
figures \ref{figure:emis}  and \ref{figure:mfp} that for temperatures $T > \EE{5}{10} \Kel$,
the neutrino physics is dominated by leptonic processes.
The dominant neutrino
production process is electron-positron pair annihilation:
\beqs
e^- + e^+ \to \nu +\bar{\nu} \, ,
\eeqs
and the neutrino mean free path length (mfp) is set by scattering off electrons and positrons:
\beqs
\nu + e^\pm  \to \nu + e^\pm\, ,
\eeqs
and similar for antineutrinos.

As the initial temperature of the fireball is high (\mbox{$T_0 \sim \EE{2}{11} \Kel$}),
we will only consider these processes in the following.

\subsection{Neutrino creation rate}
\label{sect:chi}
We express the neutrino creation rate in terms of the parameter $\chi = t_c / t_e$, where $t_c= \mE / ( V Q )$ is the cooling timescale
and  $t_e= R / c_s$ is the expansion timescale ($c_s$ is the sound speed in the fireball). This parameter bears no reference to the neutrino transparency of the plasma, which has to be taken account if
one considers cooling by neutrino emission. The emissivity of electron-positron pair annihilation is (see appendix \ref{sect:appQ}):
\beqs
Q_{\textrm{pair}} =  \EE{3.6}{33} \,  \lb T_{11} \rb^9 \ergscm \, .
\eeqs
Because $Q$ is a function of temperature, it depends on the size, energy and composition of the fireball through
equation \eqref{eq:temp}. It follows that
\beqs
\label{eq:phase:chi}
\chi = \EE{3.7}{-3} \, g^{9/4} \lb \mE_{52} \rb^{-5/4} \lb R_{6.5} \rb^{11/4} \, ,
\eeqs
where we used $c_s = c / \sqrt{3}$. For the reference values \mbox{$\mE_0 = 10^{52} \erg$} and \mbox{$R_0 =10^{6.5} \cm$}, we find
that $\chi_{\textrm{I}}  = 0.16$, which means that neutrinos are created  reasonably rapidly as compared
to the expansion timescale.
Neutrinos and antineutrinos are created in pairs by electron-positron
annihilation, so they will be present
in equal amounts.\footnote{This conclusion changes if there is an initial asymmetry between neutrinos and antineutrinos. We do not consider this here. }

A different source of neutrinos is the decay of charged pions due to photo-pion production (see appendix \ref{sect:piondecay}) 
by high-energy photons ($E_\gamma > 140 \MeV$). The energy stored in the high-energy tail of the photon distribution is
relatively small ($\sim$ 5\%). The process manifests itself as a high-energy leak,   resulting in an increased
production of electron- and muon-neutrinos with energies below $m_\mu / 2 \simeq$ 53 MeV. We will not consider this non-thermal
process in the rest of this paper.

\subsection{Optical depth}
The fireball's opacity to neutrinos is described in terms of the optical depth \mbox{$\tau = R / \lambda$}, where  $R$ is
the length scale and $\lambda$ is the mean free path (mfp). 
The mfp due to electron and positron scattering is (see appendix  \ref{sect:epscat}):
\bsub
\label{eq:op:epscat}
\beqs
\lambda^{(e)} & = & \EE{3.7}{6} \,  \lb T_{11} \rb^{-5} \cm \, , \\
\lambda^{(\mu,\tau)} & = & \EE{1.6}{7}  \, \lb T_{11} \rb^{-5}  \cm \, ,
\eeqs
\esub
where the difference originates from the fact that only electron-neutrinos  participate in charged current-interactions.
Because the mfp for neutrinos and antineutrinos is equal (assuming an equal amount
of electrons and positrons), neutrinos and antineutrinos will leave the fireball at the same time.

We consider a generic plasma
that moves from region I to II to III, Therefore, we use the value $g = g_{\textrm{I}}$ to find the optical depth
for the muon- and tau-neutrinos and  $g = g_{\textrm{II}}$ for the electron-neutrinos:
\bsub
\label{eq:phase:tau}
\beqs
\label{eq:phase:tau:e}
\tau^{(e)}  &  = &   54 \times  \lb \mE_{52} \rb^{5/4} \lb R_{6.5} \rb^{-11/4} \, , \\
\label{eq:phase:tau:mutau}
\tau^{(\mu,\tau)}  &  = &  7.4 \times  \lb \mE_{52} \rb^{5/4} \lb R_{6.5} \rb^{-11/4} \, .
\eeqs
\esub
We observe that  for reference initial conditions, \mbox{$\tau^{(e,\mu,\tau)} > 1 $} so that the fireball is opaque to neutrinos of all flavours.

\subsection{Phases of the neutrino fireball}
\label{sect:phases}
\begin{figure*}
\center
\includegraphics[width=12cm]{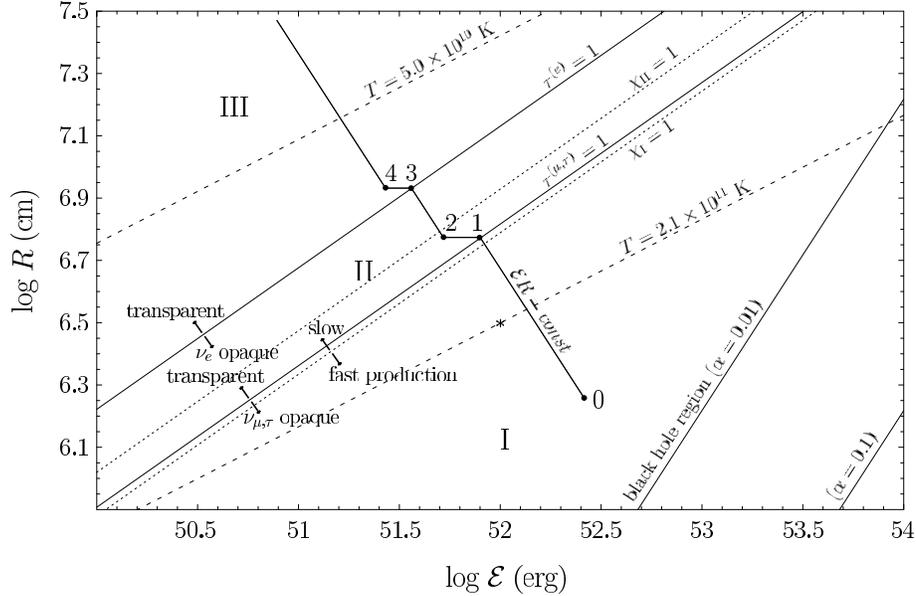} 
\caption{A parameter space plot that shows the three phases of the plasma.  The solid lines
show the $\chi_{\textrm{I}}=1$, $\chi_{\textrm{II}}=1$,  $\tau^{(\mu,\tau)}=1$ and $\tau^{(e)}=1$ contours; the dotted lines are isotemperature curves. The $*$
denotes the reference point with values given in eqs. \eqref{eq:refs}. The plotted trajectory and the points `0' to `4' are discussed in section \ref{sect:phasandevol}. The black hole lines
indicate the Schwarzschild radius as a function  of the fireball energy, assuming an initial conversion efficiency $\alpha = E^{(0)} / M_{\textrm{BH}} c^2$. }
\label{figure:phase1}
\end{figure*}
We will assume that neutrinos of some flavour decouple from the plasma instantaneously if the optical depth is one (these transitions will be smoother
in reality).
Based on equations \eqref{eq:phase:tau}, figure \ref{figure:phase1} shows how the parameter space is divided in the regions I, II and III by
the $\tau^{(\mu,\tau)}=1$ and $\tau^{(e)}=1$ contours. The dynamical evolution of a fireball through these regions will be discussed
in section \ref{sect:phasandevol}.

We observe that the region of interest has a temperature $T > \EE{5}{10} \Kel$, which justifies the fact that we only consider electron-positron pair annihilation and neutrino scattering off electrons and positrons (see figures \ref{figure:emis} and \ref{figure:mfp}).\footnote{Nuclear processes become competitive with the
leptonic processes at these temperatures if the nucleon density is approximately two orders of magnitude higher. In that case, the optical depth- and emissivity-lines in
figure \ref{figure:phase1} feature a bend at a cross-over temperature.}

The figure
also indicates the neutrino creation rate from equation \eqref{eq:phase:chi}. 
In region I this process is fast
compared to the expansion
time-scale. Together with the reverse process, it aims
towards
thermodynamic equilibrium between the neutrinos and the other
components of the plasma. The neutrinos also interact with the electrons and positrons through scattering. The interaction length of this process is 
smaller than the size of the fireball. We conclude that thermodynamic equilibrium is established rapidly, and the system will remain in equilibrium throughout its evolution.

\section{Fireball evolution}
\label{sect:evol}

\subsection{Hydrodynamics}

\label{sect:hydro}
As long as the components of the plasma
are strongly coupled (i.e. the interaction length is much smaller than the size of the system), the plasma can be described as a homogeneous sphere, in thermodynamic equilibrium with a single temperature.
The evolution will be very similar to that of a neutrinoless fireball
as described by e.g. \citet{1990ApJ...365L..55S}.
The plasma expands by radiation pressure, converting radiative energy to kinetic energy of the baryons. We assume that the
expansion is adiabatic.
We will denote the radiative energy contained in the fireball (without the
decoupled components) as $\mE$.
The energy and entropy within  a sphere of radius $R$ are
\bsub
\label{eq:consSE}
\beqs
\mE  &= & \frac{4 \pi}{3} g a R^3  T^4  \, , \\ 
 S  & = & \frac{16 \pi}{9} g a (R T)^3 \, .
\eeqs
\esub
Assuming that 
the fireball's evolution is reversible (i.e. entropy is conserved),
the temperature-radius relationship reads
\beqs
\label{eq:gRT}
g ( R T)^3 = g_0 (R_0 T_0)^3 = const \, .
\eeqs
As long as there is no change in the plasma composition, 
the following very useful scaling laws can be used to describe the evolution \citep{1990ApJ...365L..55S}:
\bsub
\beqs
\label{eq:exp:adiabatic}
\mathcal{E} R &= & \mathcal{E}_0 R_0 = const  \, , \\
\frac{\mathcal{E}}{T} & = &  \frac{\mathcal{E}_0}{T_0} = const \, .
\eeqs
\esub
If a plasma component annihilates, the temperature-radius relationship 
\eqref{eq:gRT} still holds by conservation of entropy. In the early universe, this leads to 
an increase in the photon temperature  after electron-positron
annihilation (see e.g. Weinberg \citeyear{Weinberg}), and a similar effect happens 
in the last stage of the 
neutrinoless fireball \citep{1990ApJ...365L..55S}.
By contrast,  entropy is carried away if a component decouples:
\beqs
S = S_0  - S_{\textrm{dec}} \, ,
\eeqs
where $S_{\textrm{dec}}$ is the entropy in the decoupled components. Since
$g = g_0 - g_{\textrm{dec}}$, it follows from 
eq. \eqref{eq:consSE} that the temperature-radius relationship does not change at decoupling:
\beqs
R T = R_0 T_0 \, .
\eeqs

\subsection{Neutrino decoupling bursts}
\label{sect:phasandevol}

We will discuss the hydrodynamical evolution of a fireball that starts 
in region I with a generic initial energy $\mE_0$ and size $R_0$. The trajectory is sketched in
figure \ref{figure:phase1}. As the fireball expands and cools, it will develop from
neutrino-opaque to -transparent. When this happens,  neutrinos decouple
from the plasma. 

Apart from these bursts, neutrinos are emitted continuously 
in regions where the creation rate is sufficiently high and the plasma is transparent to neutrinos. 
We will consider this in more detail in section \ref{sect:enloss}, and restrict  our discussion to an expanding
fireball with events of instantaneous energy loss here. 

Starting from the point denoted as `0' in figure \ref{figure:phase1}, the plasma expands
along a $\mathcal{E} R = \mathcal{E}_0 R_0$ line until it
reaches the $\tau^{(\mu,\tau)} = 1$ contour,
where the muon- and tau-neutrinos decouple from the plasma. 
From eq. \eqref{eq:phase:tau:mutau}, we find that the radiative energy and temperature of the plasma
just before decoupling are
\bsub
\beqs
\label{eq:ET:E1}
\mathcal{E}_{52}^{(1)} & = & 0.61 \lb \mE_{52}^{(0)} \, R^{(0)}_{6.5} \rb^{11/16} \, ,  \\
T_{11}^{(1)} &  = & 1.26  \lb \mE_{52}^{(0)} \, R^{(0)}_{6.5} \rb^{-1/16} \, .
\eeqs
\esub
The temperature of the plasma at that point depends on the initial conditions only very mildly, but
it is interesting that the temperature of the plasma at decoupling is lower if the initial energy is higher. This can be seen from figure 
\ref{figure:phase1}: for a higher $\mathcal{E}_0$, the $\mathcal{E} R = const$ line crosses the $\tau^{(\mu,\tau)} = 1$
contour at a lower temperature. The muon- and tau-neutrinos carry away $14/43 \simeq 33$\% of the available
radiative energy. This moves the fireball from point 1 to point 2.

Since the  size and temperature of the plasma are constant at decoupling, what remains of the fireball continues adiabatic expansion along a new
trajectory. The electron-neutrinos
remain in  thermal equilibrium with the plasma, which enters region II. 
The expansion continues along a $\mathcal{E} R = \mathcal{E}_2 R_2$ curve until
the plasma becomes transparent to electron-neutrinos at $\tau^{(e)} =1$ (point 3):
\bsub
\label{eq:ET:ET3}
\beqs
\mathcal{E}_{52}^{(3)}  & =  & 0.28 \lb \mE_{52}^{(0)} \, R^{(0)}_{6.5} \rb^{11/16}  \, , \\
T_{11}^{(3)} & = &  0.87  \lb \mE_{52}^{(0)} \, R^{(0)}_{6.5} \rb^{-1/16} \, .
\eeqs
\esub
At this point, the electron-neutrinos leave the plasma and carry away $7/29  \simeq 24$\%  of the energy (point 4).
When all the neutrinos are decoupled, the fireball will develop according to the standard scenario  \citep{1990ApJ...365L..55S}.

The energy that is emitted in neutrinos in the two bursts is:
\beqs
\label{E:nu:burst}
E_{52}^{(\nu \textrm{ bursts})}  =   \frac{14}{43}  \mE^{(1)}_{52} +  \frac{7}{29}   \mE^{(3)}_{52}  = 0.27
\lb \mE_{52}^{(0)} \, R^{(0)}_{6.5} \rb^{11/16} \, ,
\eeqs
which is a significant fraction of the initial radiative energy.

\subsection{Continuous neutrino cooling}
\label{sect:enloss}
In regions in the parameter space where the neutrino creation rate is high ($\chi \lesssim 1$) and (some of) the neutrinos can escape from
the plasma ($\tau \lesssim 1$), we should take neutrino cooling into account in the hydrodynamical evolution.

A plasma expanding adiabatically along a $\mE R = const$ contour, converts radiative energy to kinetic energy according
to 
\beqs
\label{eq:enloss:exp}
\left. \frac{ d \mathcal{E}}{d R}\right|_{\textrm{exp}} = - \frac{\mathcal{E}}{R} \,  .
\eeqs
To this we add the energy loss by neutrino cooling \mbox{$\Delta \mathcal{E} = - f Q V \Delta t$}, where $Q$ is the emissivity and $f$ is
the fraction of the created neutrinos that can leave the plasma. Assuming $\Delta R \simeq c_s \Delta t$, we
 find that
\beqs
\left. \frac{ d \mathcal{E}}{ d R} \right|_{\nu \textrm{ cooling} } \simeq - \frac{f Q V}{c_s}  =  -  \frac{f}{\chi } \frac{\mathcal{E}}{R} \, ,
\eeqs
where $\chi = \chi(\mE,R) $ is the creation rate as defined in section \ref{sect:chi}. The plasma evolution, including neutrino cooling, 
can then be determined from the differential equation
\beqs
\label{eq:cooling:diff}
  \frac{ d \mathcal{E}}{ d R}  =  - \lb 1+  \frac{f}{\chi} \rb  \frac{\mathcal{E}}{R} \, ,
\eeqs
so that, locally, the plasma moves along a $\mE R^{1+f / \chi} = const$ trajectory.
From eq. \eqref{eq:phase:chi}, we find that just after electron- and muon-neutrino decoupling the creation rate parameter is
\beqs
\chi_{\textrm{II}}  = 0.81 \, ,
\eeqs
independent of initial conditions. Hence the creation
rate is reasonably high in this region, where only muon- and tau-neutrinos can escape. Using the
emissivity formulae from \citet{1985ApJ...296..197M}, we find that 31\% of the neutrinos
created by electron-positron pair creation are muon- or tau-neutrinos,
so that $f = 0.31$.
Combining this with eqs. \eqref{eq:phase:chi} and \eqref{eq:cooling:diff}, we find that the plasma expands until it reaches the $\tau^{(e)}=1$ contour at
\bsub
\beqs
\mE_{52}^{(3)} & = & 0.27 \lb \mE_{52}^{(0)} \, R^{(0)}_{6.5} \rb^{11/16} \, ,
\\
T^{(3)}_{11} & = & 0.89  \lb \mE_{52}^{(0)} \, R^{(0)}_{6.5} \rb^{-1/16} \, ,
\eeqs
\esub
which is almost identical to eqs. \eqref{eq:ET:ET3}.

After electron-neutrino decoupling, neutrinos of all flavours can leave the plasma. The energy loss due to
continuous neutrino cooling in regions II and III is
\bsub
\label{E:nu:cont}
\beqs
E^{(\nu,\textrm{II})}_{52} &=& 0.027 \lb \mE_{52}^{(0)} \, R^{(0)}_{6.5} \rb^{11/16}  \, , \\
E^{(\nu,\textrm{III})}_{52} &=& 0.015 \lb \mE_{52}^{(0)} \, R^{(0)}_{6.5} \rb^{11/16}  \, .
\eeqs
\esub
The continuous energy loss component is relatively small and hardly affects the fireball evolution.
In particular, neutrino cooling is never efficient enough to prevent a hot fireball from exploding.

\section{Neutrino emission}
\label{sect:nuemission}

\subsection{Observed temperature}
For the neutrinoless fireball, it is well known that  the temperature of the observed  photon spectrum is
roughly equal to the initial temperature of the plasma $T_0$  \citep{1990ApJ...365L..55S,Goodman:1986az}.
Let us recall the thermodynamic treatment of this phenomenon \citep{Goodman:1986az}.
The number of photons in a sphere of radius $R$ depends on the temperature $T$ as
\beqs
\label{eq:hydro:Ng}
N_\gamma = \frac{2 \zeta{(3})}{\pi^2} \lb \frac{k_B T}{\hbar c} \rb^3 \lb \frac{4}{3} \pi R^3 \rb \sim 1.0 \lb \frac{ k_B}{\hbar c} \rb^3 \lb R T \rb^3 \, .
\eeqs
As long as none of the plasma components annihilates, the number of photons is constant during the evolution.
The average available energy per photon for a neutrinoless fireball is initially
\beqs
\langle E_{\gamma} \rangle^{(0)} = \frac{4}{11} \frac{E_{\textrm{tot}}}{N_{\gamma}^{(0)}} \, .
\eeqs
Since the total energy is conserved, the available energy per photon does not change during the fireball's evolution.

This conclusion is unaffected by the annihilation of electrons and positrons that occurs in the last stage of the fireball:
entropy conservation requires that the number of photons increases by a factor of 11/4. However, the
total energy is now exclusively available for the photons, so the available energy increases by the same factor.
The mean photon energy does not change during the evolution of the neutrinoless fireball  and the
observed photon spectrum  is roughly equal to the initial blackbody \citep{Goodman:1986az}, with temperature
 \citep{1990ApJ...365L..55S,1993MNRAS.263..861P}
\beqs
T_{\textrm{obs}} = \gamma T \simeq T^{(0)} \, .
\eeqs
As for photons, the mean  available energy for muon- and tau-neutrinos remains constant during the expansion from point 0 to 1, so the observed temperature 
will roughly equal the initial temperature.

For the electron-neutrinos, the situation is more subtle because energy leaves the plasma when the muon- and tau-neutrinos decouple.
Initially, the mean available energy is
\beqs
\langle E_{\nu_e} \rangle^{(0)} = \frac{7}{43} \frac{E_{\textrm{tot}}}{N_{\nu_e}^{(0)}} \, ,
\eeqs
which remains constant throughout the evolution to point 1. 
At point 2, the available energy is reduced by a factor $29/43$, but the electron-neutrinos get a larger share:
\beqs
\langle E_{\nu_e} \rangle^{(2)} = \frac{7}{29} \frac{ \frac{29}{43} E_{\textrm{tot}}}{N_{\nu_e}^{(2)}}  = \langle E_{\nu_e} \rangle^{(0)} \frac{N_{\nu_e}^{(0)}}{N_{\nu_e}^{(2)}} \, .
\eeqs
The number of neutrinos\footnote{This is similar to eq. \eqref{eq:hydro:Ng}, but for neutrinos (one flavour) the prefactor is 0.38 rather than 1.0.} in a sphere of radius 
$R$ is proportional to $(RT)^3$. 
Because $R_2 T_2 = R_1 T_1=R_0 T_0$, the mean available energy does not change when some plasma components decouple. We conclude that
 the observed temperature of the electron-neutrino spectrum is also approximately equal to $T^{(0)}$.

\subsection{Energy}

\begin{table}
\center
\begin{tabular}{ | c | c  | c  c  c | }
\hline
& $E^{(\nu \, \textrm{tot})}_{52} $ & $\nu_e$ \, \,  : & $\nu_\mu $ \, \, : & $\nu_\tau$  \\
\hline
$\nu_{\mu,\tau}$  dec. &  $0.20    \times \xi_0 $& 0 & 0.5 & 0.5 \\
$\nu_{e}$  dec. &  $0.070  \times \xi_0 $ & 1 & 0 & 0 \\
cont., II &  $0.024  \times  \xi_0 $ & 0 & 0.5 & 0.5 \\
cont., III & $ 0.013  \times \xi_0 $ & 0.69 & 0.15 & 0.15 \\
\hline
total &  $ 0.31  \times \xi_0 $  & 0.26 & 0.37 & 0.37 \\
\hline
\end{tabular}
\caption{The total energy that is emitted in neutrinos in various
stages. Here `dec.' stands for decoupling bursts, `cont.' for continuous emission. The symbol $\nu$ means `neutrino and antineutrino' in the above,
and  $ \xi_0  :=  (  E^{(0)}_{52} \, r^{(0)}_{6.5}  )^{11/16} $}
\label{table:nuemission}
\end{table}

The evolution of a fireball with neutrinos
is described in section \ref{sect:evol}. Using the results
obtained in eqs.
\eqref{E:nu:burst} and \eqref{E:nu:cont}, we summarize the neutrino
emission in table \ref{table:nuemission}.
Neutrinos and
antineutrinos are emitted in equal amounts and share the energy quoted in the
table. The total energy that is emitted in neutrinos equals
\beqs
\label{eq:res:E:tot}
E^{(\nu, \, \textrm{tot})} =\EE{3.1}{51} \erg \times \lb E_{52}^{(0)} R_{6.5}^{(0)} \rb^{11/16} \, .
\eeqs
The mean neutrino energy follows directly
from the initial temperature:
\beqs
\label{eq:res:E:temp}
\langle E_\nu \rangle = 3.15 k_B T^{(0)} = 56 \MeV \times \lb E_{52}^{(0)} \rb^{1/4} \lb R_{6.5}^{(0)} \rb^{-3/4} \, .
\eeqs

\subsection{Time spread}
The neutrinos are emitted in two decoupling bursts as well as continuously. As is clear from
figure \ref{figure:phase1}, the fireball has not expanded much in between the two decoupling
events: $R^{(1)} - R^{(3)} \sim R^{(1)}$,  implying that the various components of neutrino
emission overlap in time. During the expansion of the fireball, $\gamma / R = const$ 
\citep{1993MNRAS.263..861P}, so that (due to Lorentz contraction) an inertial observer sees a time  spread corresponding
to $R^{(0)}/c$. Hence
\beqs
\Delta t_{\textrm{obs}} = \frac{R^{(0)}}{c} \sim 0.1\,  \textrm{ms} \times R^{(0)}_{6.5} \, ,
\eeqs
which is much smaller than the typical time spread for supernova neutrinos that originate
from relatively slow deleptonization processes.

Dispersion effects on the way to Earth introduce an additional smearing:
\beqs
\Delta t_{\textrm{disp}} & = &  \frac{D}{c} \lb \frac{1}{\beta} -1 \rb \\
\nn & = &  0.6 \,  \textrm{ms}
\times \lb \frac{m_\nu}{0.1 \,  \textrm{eV}} \rb^2 \lb \frac{E_\nu}{56 \,  \textrm{MeV}} \rb^{-2} 
\lb \frac{D}{4 \,  \textrm{Mpc}} \rb^2 \, .
\eeqs
For a robust analysis, this time spread should be averaged over a thermal distribution.

\subsection{Applications}
The detectability of a neutrino source as described in this paper was studied by \citet{Halzen:2002pg} (see also \citet{Halzen:1995ex,Halzen:1996qw}). The detection is based on the charged current interaction $\bar{\nu}_e + p \to n + e^+$ and the
subsequent \v Cerenkov radiation that is emitted by the positron. 
An analysis based on \citet{Halzen:1996qw} shows that
detection could be feasible for sources within a few Mpc for a  low-background neutrino
telescope.\footnote{This is a rough signal-over-noise estimate. In particular, it assumes that there is no directional information available for triggering or reconstruction. The observational time window is 0.3 ms.}
This limits potential sources to our local cluster.

In the context of supernova dynamics,  it has been proposed that delayed neutrino emission
could revive a stalled supernova shock \citep{Bethe:1984ux}. Matter that is surrounding some
central, heavy object can escape if the internal energy exceeds the gravitational energy:
\beqs
E_{\textrm{int}} > E_{\textrm{grav}} = \frac{G M}{D} \, ,
\eeqs
where $M$ is the mass of the central object and $D$ the distance of the matter 
to the central object.

This material can be heated by neutrinos. We assume that the matter consists of
nucleons, but
for heavier nuclei similar processes can occur.
Neglecting loss terms, the total energy that can be deposited by neutrinos
from the central object equals
\beqs
\Delta E = N_A \sigma \frac{E^{(\nu, \, \textrm{tot})}}{4 \pi D^2} \, ,
\eeqs
where the cross section for neutrino capture on nucleons reads
(the relevant processes and cross-sections can be found in appendix \ref{sect:appmfp}):
\beqs
\sigma \sim 10^{-43} \cm^2 \times \left< \frac{E_\nu^2}{1 \MeV^2} \right> \, .
\eeqs
Following \citet{Bethe:1984ux}, we use the values
\mbox{$D = 150 \, \textrm{km}$} and \mbox{$M = 1.6 M_{\odot} = \EE{3}{54} / c^2 \erg$}. With the expressions
\eqref{eq:res:E:tot} and \eqref{eq:res:E:temp} for the neutrino flux found in this paper, we find that
\bsub
\beqs
E_{\textrm{grav}} & \sim &  \EE{2}{19} \erg  \textrm{ g}^{-1} \, , \\
\Delta E & \sim &  \EE{2}{20} \erg  \textrm{ g}^{-1} \, .
\eeqs
\esub
We conclude that the energy released in neutrinos by the hot fireball considered here
is sufficient to release material at a typical distance $D \sim 150$ km from the gravitational pull of
a $1.6 M_{\odot}$ object.

\section{Conclusions}
\label{sect:conclusions}

In this paper, we have described the physics of neutrinos in a hot
fireball environment. We find that the dominant neutrino processes are leptonic:
neutrino creation by electron-positron annihilation and neutrino scattering
off electrons and positrons.

 For general initial conditions\footnote{These conclusions apply to fireballs that starts in the neutrino-opaque region 
 that we denoted as region I. This is the case if
\beqs
 \lb E_{0,52} \rb^{-5/4} \lb R_{0,6.5} \rb^{11/4} \lesssim 5
\eeqs
}, the fireball plasma is initially neutrino-opaque and the rate of neutrino creation is reasonably
 high. The
neutrinos are in thermal equilibrium with the other components of the plasma and follow the hydrodynamical evolution of the fireball.
In this evolution, the muon- and tau-neutrino decouple first, followed by
the electron-neutrinos. Besides these bursts, the fireball emits neutrinos
continuously in regions where it is neutrino-opaque and the creation rate is
high. The effect on the evolution of the fireball and on the neutrino emission is small.

The energy spectrum of the emitted neutrinos will be approximately thermal with a temperature equal to the initial temperature of the
fireball, i.e. $\langle E_\nu \rangle \sim 60$ MeV.
The total energy that is emitted in (anti)neutrinos is 
\beqs
E^{(\nu, \, \textrm{tot})} =\EE{3.1}{51} \erg \times \lb  E^{(0)}_{52} \, R^{(0)}_{6.5} \rb^{11/16} \, .
\eeqs
A sizable fraction of the total fireball's energy is converted into neutrinos,
and this fraction is not very sensitive to initial conditions.
 The rather limited detection possibility is
mainly due to the isotropic outflow of the neutrinos, as opposed to the
observed high-energy gamma-rays that originate in ultra-relativistic  beamed jets
in a later
stage of the GRB. If the neutrinos were focused by some mechanism,
detection of sources much further away could be possible.
On the other hand, fewer sources will be detected since the 
outflow needs to be directed towards the Earth.

We have found that our initial concern that neutrino emission 
might prevent the production of powerful explosions from fireballs is
not justified.
The physical reason for this is that for most of the parameter space where
neutrino production is fast enough to cool the fireball, the fireball 
shields itself from cooling by being opaque to those same neutrinos.
However, there may well be another snag when one considers the formation
of the fireball: this requires a heating mechanism, and at the start
of the heating one necessarily approaches the safe zone in the lower right 
half of figure 1 from the left. Therefore, unless the heating occurs on
a timescale close to the dynamical time the evolution track towards high 
energy may well get stuck in the cooling zone, causing loss of all heating
energy into neutrinos. Given that the dynamical timescale is probably the
fastest thinkable heating time, it is quite possible that neutrino
cooling can prevent high-energy fireballs from forming.

\section*{acknowledgments}
We thank Francis Halzen, Maarten de Jong and Eli Waxman for useful discussions.

\appendix
\section{Neutrino emitting processes}
\subsection{Direct neutrino production}
\label{sect:appQ}
There is extensive literature on neutrino emitting processes in an electroweak plasma or in a nuclear environment. We refer the reader to 
\citet{Dicus:1972yr,Braaten:1993jw,Bruenn:1985en,Dutta1,1989ApJ...339..354I,1985ApJ...296..197M,Ratkovic:2003td,Baiko:1998jq,1979ApJ...232..541F,Lattimer:1991ib,Qian:1996xt} and further
references therein for a broader overview on the subject. In the hot fireball environment, the most important
processes are:
\begin{figure}
\center
\includegraphics[width=7cm]{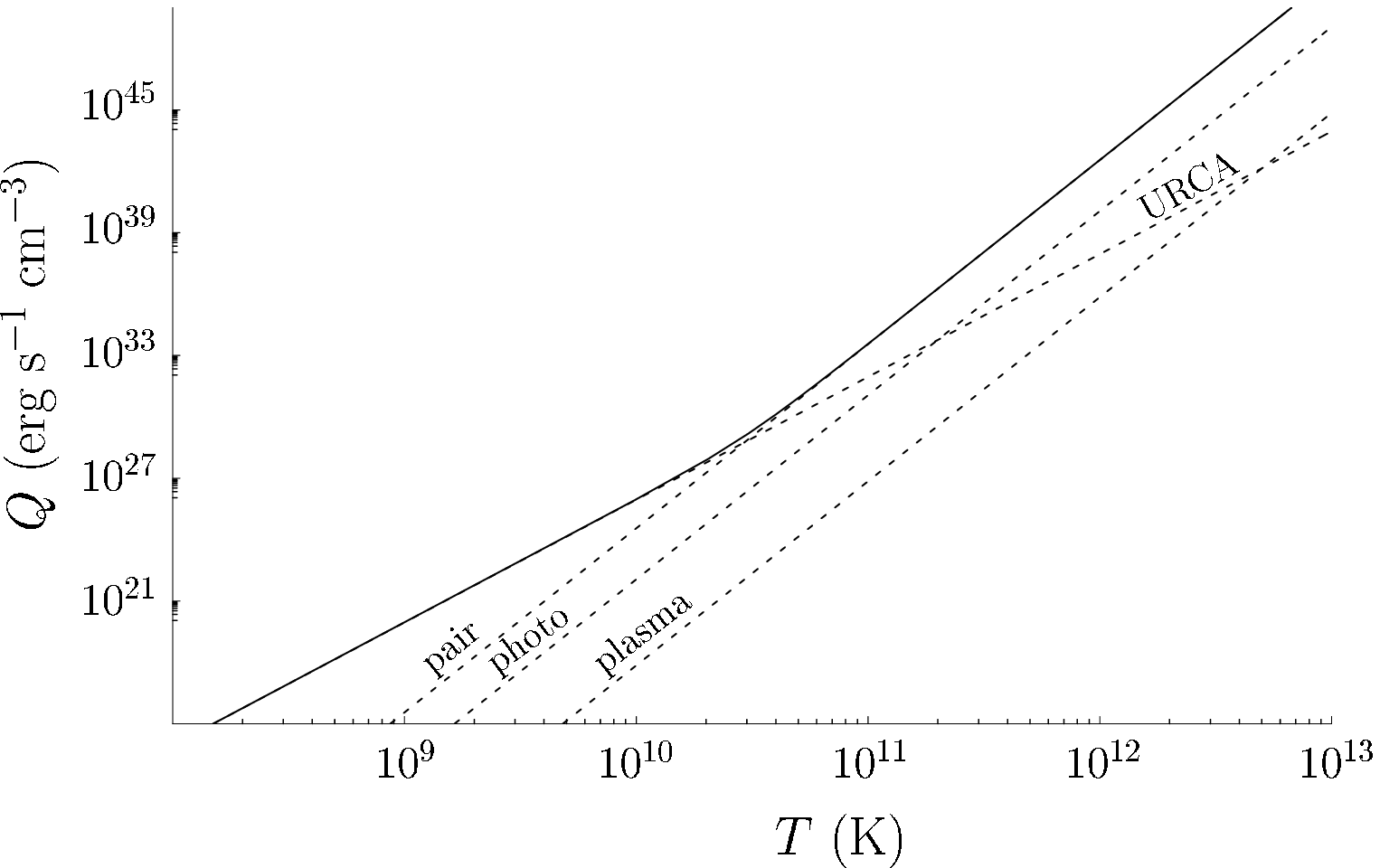}
\caption{Neutrino emissivity of the plasma as a function of temperature. The dashed lines
show the individual contributions, the solid line shows the total emissivity. We used a baryon
density $\rho = 10^8$ gr cm$^{-3}$.}
\label{figure:emis}
\end{figure}
\beqs
\nn
\begin{array}{rlrcl}
\bullet  &  \textrm{photo-neutrino process:} & e^\pm + \g & \to & e^\pm + \nu_i + \bar{\nu}_i \, ; \\
\bullet  & \textrm{plasma process:} & \g & \to & \nu_i + \bar{\nu}_i \, ; \\
\bullet  & \textrm{pair annihilation:} & e^- + e^+ & \to & \nu_i + \bar{\nu}_i \, ; \\
\bullet  & \textrm{electron capture:} & e^- + p & \to & n + \nu_e \, ; \\ 
\bullet  & \textrm{positron capture:} & e^+ + n & \to & p + \bar{\nu}_e \, .
\end{array}
\eeqs
The last two processes constitute the non-degenerate URCA process, which is the dominant
nuclear neutrino emitting process for low nucleon densities.
Neutron decay is too slow to play a role of importance if the neutrons are non-degenerate. 

We use the following total (i.e. adding all neutrino flavours) emissivities for the photoneutrino \citep{Dutta1}, plasma \citep{Ratkovic:2003td}, pair annihilation 
\citep{1989ApJ...339..354I} and  non-degenerate URCA  \citep{Qian:1996xt} processes:
\bsub
\beqs
Q_{\textrm{photo}} & = &  \EE{1.1}{31} \, \lb T_{11} \rb^9 \ergscm \, ; \\
Q_{\textrm{plasma}} & = & \EE{7.1}{26} \, \lb T_{11} \rb^9 \ergscm \, ; \\
\label{eq:Q:pair} Q_{\textrm{pair}} &  = &  \EE{3.6}{33} \,  \lb T_{11} \rb^9 \ergscm \, ; \\
\label{eq:Q:ndURCA} Q_\textrm{URCA}  &= &  \EE{9.0}{31} \, \lb T_{11} \rb^6
\lb \rho_{B,8} \rb \ergscm \, ,
\eeqs
\esub
where $\rho_B = \rho_{B,8} \times 10^8 \, \textrm{gr cm}^{-3}$.
These emissivities are plotted as a function of temperature in figure \ref{figure:emis}.
The emissivity of both the photoneutrino and the plasma process 
is several orders of magnitude lower than that of $e^- e^+$ pair annihilation,
which is in keeping with similar comparisons in the literature \citep{1989ApJ...339..354I,Prakash:2004wv,Raffelt:1996wa}.  

Electron-positron pair annihilation and non-degenerate URCA have a different
scaling behaviour with temperature, and the URCA process depends on baryon density. 
For the environment considered in this study,  we conclude that pair annihilation is the dominant process.

\subsection{Neutrinos from pion decay}
\label{sect:piondecay}
Another source of neutrinos  is the decay of  charged pions:
\beqs
\nn
\begin{array}{rlrcl}
\bullet  &  \textrm{pion decay:} & \pi^-  & \to & \mu^- + \bar{\nu}_\mu \\
& &  &\to &  e^- + \bar{\nu}_e + \nu_\mu + \bar{\nu}_\mu \, , \\
\end{array}
\eeqs
and the charge-conjugate process for  $\pi^+$ decay.
The pions originate from photopion production
or nucleon-nucleon collisions:
\beqs
\nn
\begin{array}{rlrcl}
\bullet  &  \textrm{photo-pion production:} & \gamma + n  & \to & p + \pi^- \, ; \\
 & & \gamma + p & \to & n + \pi^+ \, ; \\
\bullet  & \textrm{N-N collisions:} &  n + p & \to &  p + p + \pi^- \, ; \\
  & &  p + p & \to &  n + p + \pi^+ \, .
\end{array}
\eeqs
The cross-section of pion production in nucleon-nucleon collisions ($\sigma \sim \EE{3}{-26} \cm^2$, see e.g. \citet{Bahcall:2000sa}) is larger
than that of the photo-pion process  ($\sigma \sim 10^{-28} \cm^2$, see e.g. \citet{Mucke:1998mk}), but the photon density in the plasma is almost four orders of magnitude higher.
This  means that photo-pion production is the dominant pion creating process.

Pion production can only occur at energies above the 
pion mass threshold  $E_t \sim 140 \MeV$. This implies that only photons in the high-energy tail of the distribution (constituting less than 5\% of the
total energy in photons) can create pions. Most of the pions are created at threshold, and decay into muon- and electron-
(anti)neutrinos with energies below $m_\mu/2 \simeq 53$ MeV. The energy spectrum of the various neutrino types is different, but the mean energies are in the range of 31 to 37 MeV.

\section{Neutrino absorption and scattering processes}
\label{sect:appmfp}
\begin{figure}
\center
\includegraphics[width=7cm]{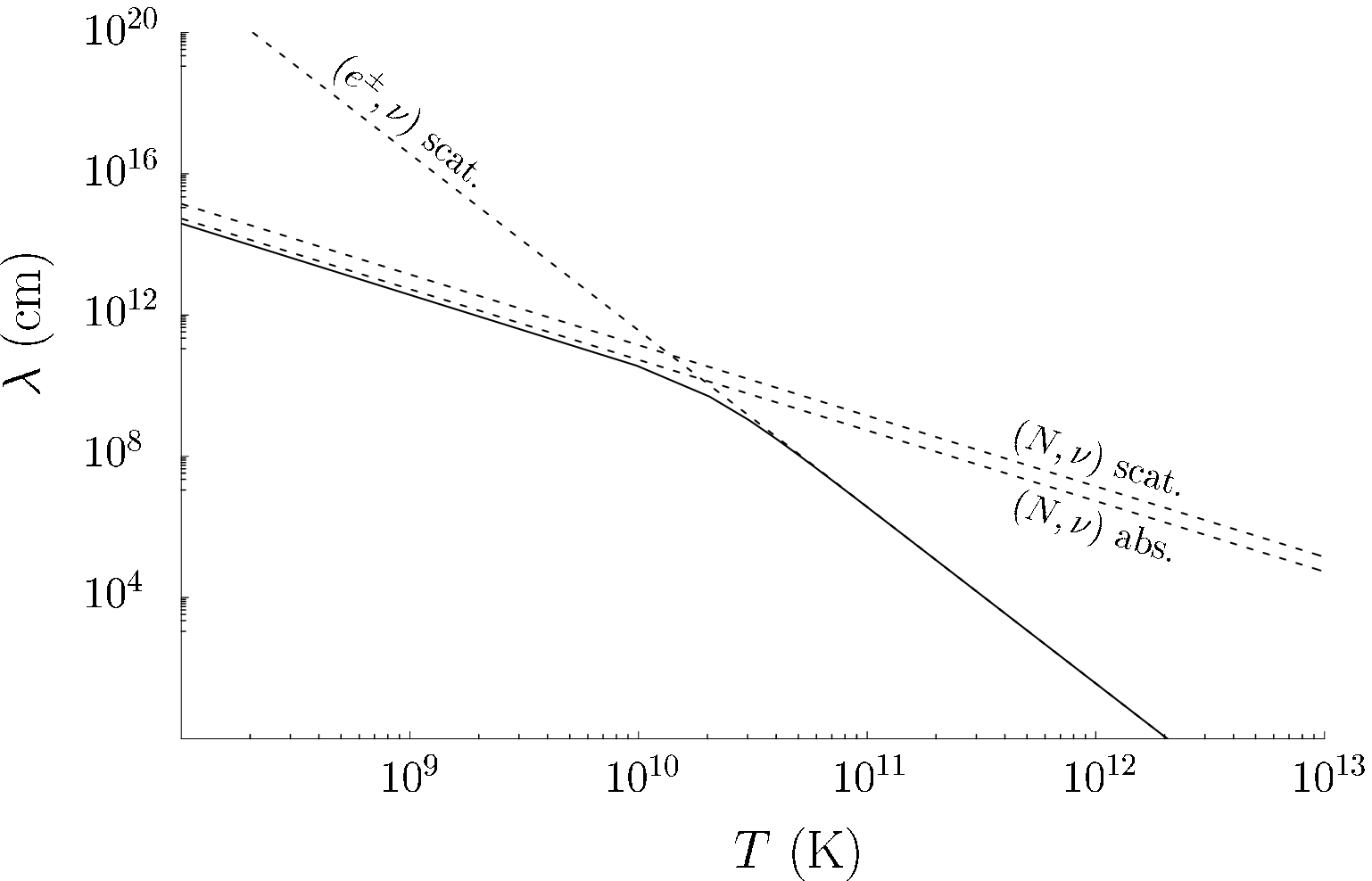} 
\hspace{0.5cm}
\includegraphics[width=7cm]{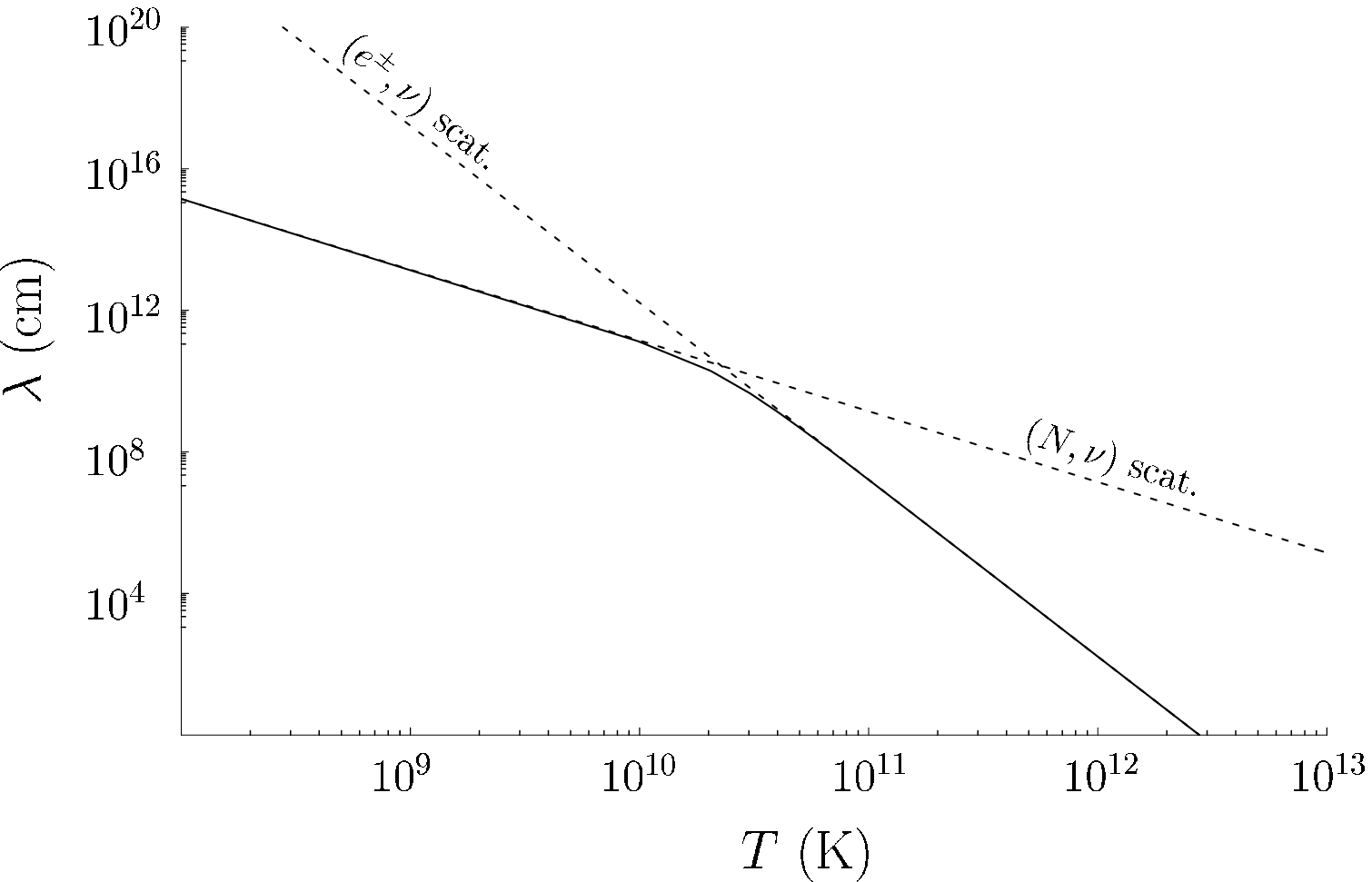}
\caption{Neutrino mean free path lengths (cm) as a function of temperature (K). The first panel
applies to electron neutrinos, the second to muon- and tau-neutrinos.
The dashed lines show the individual contributions to the mean free path length, the solid line
is the combined mean free path length. We used the value $Y_e=0.5$ for nucleon scattering. 
The graphs for the corresponding antineutrinos are virtually
identical.}
\label{figure:mfp}
\end{figure}

We summarize the cross-section formulae for the following processes:
\beqs
\nn
\begin{array}{rlrcll}
\bullet  & \textrm{$e^\pm$ -- neutrino scattering:} & \nu_i + e^\pm & \to & \nu_i + e^\pm  \, ;  \\
\bullet  & \textrm{nucleon -- neutrino scattering:}  & \nu_i + N & \to &  \nu_i + N  \, ; \\
\bullet  & \textrm{electron-neutrino capture:} & \nu_e + n & \to &  p + e^-  \, ; \\
\bullet  & \textrm{electron-antineutrino capture:} & \bar{\nu}_e + p & \to &  n + e^+  \, ,
\end{array}
\eeqs
where we assume that all the particles are non-degenerate. The result, in terms of the mean free path length (mfp), is plotted in figure
 \ref{figure:mfp}. We use number densities  $n_{e^-} = n_{e^+} = \EE{1.4}{35} \cm^{-3} $ and $n_{B} =  \EE{5}{31} \cm^{-3} $.

From figure \ref{figure:mfp}, we conclude that the neutrino mfp in the fireball is determined by scattering off electrons and positrons.

\subsection{Electron and positron scattering}
\label{sect:epscat}
The cross-section\footnote{The vacuum cross section 
scales as $T$ via the neutrino energy \citep{'tHooft:1971ht,Sehgal:1974wz}, but it is important to realize that we consider plasma cross-sections.
The role of the electron mass in the vacuum cross-section  is taken by the thermal energy, which
leads to an increase by a factor \mbox{$3.15 k_B T / m_e$}. For a temperature $T_*$, this is two orders
of magnitude.}  for neutrino scattering off electrons in a plasma is \citep{1975ApJ...201..467T}: 
\beqs
\label{eq:scat:e}
\sigma  =   \frac{3 G_F^2 \hbar^2 c^2}{2 \pi}  \lb \lb  c_V +c_A \rb^2 + \frac{\lb c_V- c_A \rb^2}{3}   \rb \lb k_B T \rb  E_\nu \, ,
\eeqs
where $G_F^2 \hbar^2 c^2 = \EE{5.3}{-44} \cm^2 \MeV^{-2}$ and
\beqs
c_V = 1/2 + 2 \sin^2 \theta_w   \, , \quad   c_A= 1/2\, , \quad \sin^2 \theta_W = 0.22 \, .
\eeqs
We average over a thermal neutrino distribution by replacing $E_\nu \to \langle E_\nu \rangle = 3.15 \, k_B T$.
The formula as it stands applies to electron-neutrinos, which interact with electrons through
both the charged and neutral current.
For other neutrinos, one should make the following substitutions  \citep{1975ApJ...201..467T}:
\beqs
\begin{array}{r r c l r c l}
\nu_\mu, \nu_\tau: &  c_A  & \to & c_A - 1 \, ,  &  c_V & \to & c_V -1 ;\\
\bar{\nu}_e:& c_A & \to & - c_A  \, ,  & c_V &  \to & c_V  ;\\
\bar{\nu}_\mu, \bar{\nu}_\tau:& c_A & \to &   1-c_A \, , &  c_V  & \to &  c_V-1 .\\
\end{array}
\eeqs
For muon- and tau-neutrinos, this accounts for the fact that
these only have a neutral interaction with electrons.
The cross-section for neutrino -- positron scattering is equal to the cross-section for
the scattering of the corresponding antineutrino off an electron.
If the electron and positron densities are equal, these processes can be combined as follows:
\beqs
\label{eq:signa:epscat}
\nn \sigma  \lb \textrm{$\nu_i, e^\pm$} \rb  &= & \sigma  \lb \textrm{$\nu_i, e^-$} \rb +  \sigma  \lb \textrm{$\nu_i, e^+$} \rb \\
&= & \sigma  \lb \textrm{$\nu_i, e^-$} \rb +  \sigma  \lb \textrm{$\bar{\nu}_i, e^-$} \rb \, ,
\eeqs
and the mean free path length due to combined electron-positron scattering follows from
\beqs
\lambda^{-1} \lb \textrm{$\nu_i, e^\pm$} \rb  = \sigma  \lb \textrm{$\nu_i, e^\pm$} \rb  n_{e^-} \, .
\eeqs
Because the electron and positron density scales as $T^3$, the mean free path length is proportional
to $T^{-5}$.

\subsection{Nucleon scattering}
Neutrino -- nucleon scattering is independent
of neutrino flavour because the interaction is neutral. From \citet{Raffelt:1996wa}: 
\beqs
\sigma   =   \frac{ G_F^2 \hbar^2 c^2}{\pi}  \lb C_V^2 + 3 \, C_A^2 \rb   {E_\nu}^2 \, ,
\eeqs
where we understand that  $E_\nu^2 \to  \langle E_\nu^2 \rangle = 12.9 \lb  k_B T \rb^2$.
Neutrino -- proton and neutrino -- neutron scattering have slightly different cross-sections because of
different strong interaction form factors\footnote{We use
the values $C_V^2 = 0.0012$  $(0.25)$ and $C_A^2 = 0.47$ $(0.33)$ for protons (neutrons) \citep{Raffelt:1996wa}.}
$C_V$ and $C_A$.  We average the cross-section by assuming an equal amount of neutrons
and protons ($Y_e = 0.5$):
\beqs
\sigma  \lb \nu_i, N \rb = \sigma  \lb \textrm{$\nu_i, p$} \rb + \sigma  \lb \textrm{$\nu_i, n$}  \rb  \, , 
\eeqs
and compute the mean free path from
\beqs
\lambda^{-1}   \lb \nu_i, N \rb = \sigma  \lb \nu_i, N \rb  \lb 0.5 n_B \rb \, .
\eeqs
The baryon density is independent of temperature\footnote{The baryon density does not scale with temperature in a dynamical way. Indirectly, the quantities are related
by the requirement that there should be 1 TeV per baryon: a higher temperature permits a higher baryon density.}
 so that the mean free path length is proportional to $T^{-2}$.

\subsection{Nucleon absorption}
Electron-neutrinos and -antineutrinos can be absorbed by neutrons and protons through  the charged 
interaction. The cross-section  is \citep{1975ApJ...201..467T}
\begin{subequations}
\beqs
\sigma & = &  \frac{ G_F^2 \hbar^2 c^2}{\pi}  
\lb 3 \alpha^2 +1  \rb { E_\nu }^2 g(E_\nu) \, , \\
 g(E_\nu) & = & \lb 1 \pm \frac{Q}{E_\nu} \rb \lb 1 \pm 2 \frac{Q}{E_\nu} + \frac{Q^2 - (\pm m_e^2)}{E_\nu^2} \rb^{1/2} \, ,
\eeqs
\end{subequations}
where $\alpha=-1.26$ is the nuclear axial coupling coefficient and $Q= 1.3 \MeV$ is the neutron-proton mass difference.
The positive sign applies to neutrino capture on neutrons, the negative sign to antineutrino capture on protons. 
We do not average cross sections here, because each process is specific to either electron-neutrinos or electron-antineutrinos.
Averaging over a thermal neutrino distribution is understood as in the nucleon scattering cross section, and (up to
small corrections due to the energy dependence of the function $g$)  the
mean free path length is proportional to  $T^{-2}$. 

\label{lastpage}

\end{document}